\documentclass[twocolumn,showpacs,preprintnumbers,amsmath,amssymb]{revtex4}
\usepackage{graphicx}
\usepackage{dcolumn}
\usepackage{bm}
\usepackage{subfigure}
\usepackage[colorlinks,linkcolor=blue,anchorcolor=blue,citecolor=blue]{hyperref}

\begin{document}

\title{Quantum $K$-nearest neighbor classification algorithm based on Hamming distance}
\author{Jing Li$^{1}$}
\author{Song Lin$^{2}$}
\altaffiliation{Corresponding author. Email address: lins95@gmail.com}
\author{Kai Yu$^{1}$}
\author{Gong-De Guo$^{1}$}
\affiliation{
$^{1}$College of Mathematics and Informatics, Fujian Normal University, Fuzhou 350117, China\\
$^{2}$Digital Fujian Internet-of-Things Laboratory of Environmental Monitoring, Fujian Normal University, Fuzhou 350117, China}
\date{\today}

\begin{abstract}
$K$-nearest neighbor classification algorithm is one of the most basic algorithms in machine learning, which determines the sample's category by the similarity between samples. In this paper, we propose a quantum $K$-nearest neighbor classification algorithm with Hamming distance. In this algorithm, quantum computation is firstly utilized to obtain Hamming distance in parallel. Then, a core sub-algorithm for searching the minimum of unordered integer sequence is presented to find out the minimum distance. Based on these two sub-algorithms, the whole quantum frame of $K$-nearest neighbor classification algorithm is presented. At last, it is shown that the proposed algorithm can achieve a quadratical speedup by analyzing its time complexity briefly.
\end{abstract}

\pacs{03.67.Dd, 03.65.Ta, 03.67.Hk}
\keywords{quantum machine learning, $K$-nearest neighbor classification, quantum algorithm}
\maketitle

\section{Introduction}
\par Recently, with the development of quantum mechanics and information science, quantum information, the product of the combination of these two disciplines, has gradually attracted people's attention. One of the most popular issues is quantum machine learning (QML)~\cite{ref1}. Since the proposal of quantum linear system algorithm by Harrow et al.~\cite{ref2}, a series of quantum algorithms have been presented to solve various machine learning tasks, such as quantum dimensionality reduction algorithm~\cite{ref3,ref4,ref5}, quantum regression algorithm~\cite{ref6,ref7}, quantum association rules mining~\cite{ref8}, quantum decision tree classifier~\cite{ref9}, quantum support vector machine~\cite{ref10,ref11,ref12}, quantum nearest neighbor classification algorithm~\cite{ref13,ref14,ref15,ref16,ref17,ref18}, and so on. They combine classical algorithms with quantum computation, and are shown to achieve significant speedup over their classical counterparts.

\par As one of the most common algorithms in machine learning, classification has been widely applied in image recognition~\cite{ref17} and text categorization~\cite{ref19}. Typically, $K$-nearest neighbor (KNN) classification algorithm determines the testing sample's category based on the metric of distance with excellent performance, and has been combined with quantum computation as well, arousing widespread interest of scholars. In 2011, Lloyd et al. presented a quantum method to calculate Euclidean distance~\cite{ref14} with the help of swap test~\cite{ref20}. Based on that, Wiebe et al. proposed a quantum version of nearest-centroid classification algorithm~\cite{ref15} where Grover's search algorithm~\cite{ref21}, in the form of the D\"{u}rr H\o yer minimization algorithm~\cite{ref22} was utilized to find the closest cluster. However, for non-numerical data, it doesn't make sense to compute the Euclidean distance, while Hamming distance is significative and can be obtained easily. In terms of that, Ruan et al. proposed a quantum KNN algorithm based on Hamming distance~\cite{ref16}. However, in their algorithm, besides $K$, a new parameter $t$ is added, which is used to find out the training samples from whom the distance to the test sample is less than $t$, so that there are two parameters to set and optimize, increasing the amount of computation. Moreover, a key step of KNN algorithm, searching the $K$ nearest neighbors, is ignored.

\par Further studying on these problems, we propose a whole quantum KNN classification algorithm based on Hamming distance, which is shown to be quadratically faster than its classical counterpart when the sample vectors lie in a low-dimensional feature space. In the proposed algorithm, a core sub-algorithm is put forward to search the minimum distance which is more efficient and more applicable to integer data than  D\"{u}rr H\o yer minimization algorithm~\cite{ref22} which is utilized commonly in quantum machine learning algorithms~\cite{ref15,ref17}. Moreover, we present the whole quantum frame of KNN classification algorithm where the testing sample's category can be obtained clearly.

\par The rest of this paper is organized as follows. In Sec.~\ref{sec:2}, we review the classical KNN classifier. Sec.~\ref{sec:3} presents two quantum sub-algorithms, calculating Hamming distance and searching the minimum of unordered integer sequence. Sec.~\ref{sec:4} gives the whole quantum frame of KNN classification algorithm. Conclusions are given in the last section.

\section{\label{sec:2}Preliminaries}

  \par In this section, we briefly review the basic idea and main processes of the classical KNN classification algorithm~\cite{ref19}.
  \par
\begin{figure}
    \centering
    \includegraphics[width=0.24\textwidth]{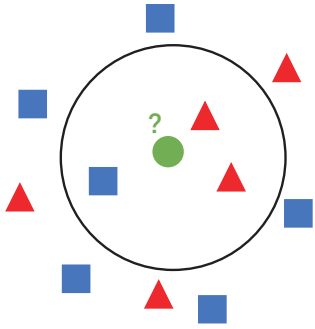}
    \caption{The illustration of KNN algorithm.}
    \label{Fig:1}
  \end{figure}

 KNN is one of the most common classification algorithms of supervised learning, where the testing sample is classified according to the similarity between it and training samples. For example, as shown in Fig. \ref{Fig:1}, there are two classes, blue square and red triangle. The task is to determine what the green circle whose class is unknown belongs to. At first, the distance metric is utilized to calculate similarities between the green circle and other samples. It's easy to see that the $K$ nearest neighbors include two red triangles and one blue square when $K = 3$. Finally, based on ``majority voting'' principle, the green circle is classified to the category of red triangle.
   \par The general processes of classical KNN classifier can be summarized as follows.
   \par (1) Compute the similarity between the testing sample (unclassified sample) and each training sample.
   \par (2) Find out $K$ nearest neighbors of the testing sample.
   \par (3) Count categories of $K$ nearest neighbors, then assign the most frequent category to the testing sample.

  \par Obviously, the runtime of KNN classification algorithm is dominated by the computation of distance which should be implemented $M$ times. Thus, its time complexity is $O\left( M \right)$.
  \par There are many ways to calculate similarity. One of the most common metrics is Hamming distance. It can be applied to KNN classification algorithm to classify non-numerical data points efficiently, which is concerned in this paper. Given the testing sample vector $\overrightarrow{x}=\left(x_{1}, x_{2}, \cdots, x_{N} \right)^{T}$ ($x_{j} \in \left\{0,1 \right\}$; $j=1,2,\cdots ,N$), whose class is unknown, and training sample vectors  $\overrightarrow{v_{i}} ={\left(v_{i1}, v_{i2}, \cdots, v_{iN} \right)}^{T}$ ($v_{ij} \in \left\{0,1 \right\}$; $j=1,2,\cdots ,N$) with class $c_{i}$, where $i=1,2,\cdots ,M$; $c_{i}\in \left\{0,1,\cdots ,L \right\}$. Hamming distance between $\overrightarrow{x}$ and $\overrightarrow{v_{i}}$ is given by
  \begin{equation}
  d_{i}=\left | \overrightarrow{x}-\overrightarrow{v_{i}}\right |=\sum_{j=1}^{N}\left ( x_{j} \oplus v_{ij}\right ),
  \label{eq:1}
  \end{equation}
  which shows the difference of two bit vectors. For example, the Hamming distance between 01101 and 11001 is 2.

Ruan et al. presented a quantum KNN classification algorithm for implementing this algorithm based on the metric of Hamming distance~\cite{ref16}. In their algorithm, a new parameter $t$ is introduced to help finding out the $K$ nearest neighbors. Specifically, if the Hamming distance between a training sample and the test sample is less than $t$, it is considered to be one of the $K$ nearest neighbors. Obviously, the value of $t$ is difficult to determine because there is no direct correlation between $t$ and $K$. In addition, generally speaking, the value of $K$ is much less than the number of training samples $M$. Then if the $K$ nearest neighbors are obtained by performing a projection measurement on the quantum register directly (mentioned in Sec. 3.2 of Ref.~\cite{ref16}), the success probability $\frac{K}{M}$ may be very small. Therefore, this problem should also be considered in the practical implementation of the algorithm presented in Ref.~\cite{ref16}. In the next section, we will give possible solutions to these problems.
\section{\label{sec:3} Two quantum sub-algorithms}

\par From the classical frame of KNN classification algorithm, it is evident that there are two key steps, i.e. steps (1) and (2). So, before presenting the whole quantum frame of KNN classifier, two corresponding quantum sub-algorithms are put forward in this section. The first quantum sub-algorithm is computing Hamming distance, which is described in Sec.~\ref{subsec:3.1}. The second sub-algorithm in Sec.~\ref{subsec:3.2} is a quantum method for searching the minimum of unordered integer sequence. In Sec.~\ref{subsec:3.3}, we perform runtime analysis on the two sub-algorithms respectively.
\subsection{ \label{subsec:3.1}Quantum method for computing Hamming distance}

  \par Computing similarity is an important subprogram in classification algorithms. For the classification of non-numerical data, Hamming distance is one of the popular ways to calculate similarity. Here, we describe a quantum method to calculate Hamming distance between $\overrightarrow{x}$ and $\overrightarrow{v_{i}}$ in parallel.
  \par A1:   Prepare the superposition state
  \begin{equation}
  \left | \phi _{1}\right \rangle=\frac{1}{\sqrt{M}}\sum_{i=1}^{M}\left | i \right \rangle\left | \bm{v_{i}}\right \rangle \left| \bm{x} \right\rangle .
  \label{eq:2}
  \end{equation}
  where $ | \bm{x} \rangle= | x_{1} x_{2} \cdots x_{N}\rangle$$=$$| x_{1}\rangle |x_{2}\rangle \cdots |x_{N} \rangle$ $=$ $  | x_{1}\rangle \otimes|x_{2}\rangle\otimes\cdots \otimes$ $|x_{N}\rangle$ and $\left | \bm{v_{i}}\right \rangle = \left | v_{i1}v_{i2} \cdots v_{iN}\right \rangle$ respectively.
  \par Considering a binary training data set $\overrightarrow{v_{i}}$, we can use a decimal value $\bm{v_i}$ to represent the vector, where $\bm{v_i}=v_{i1} 2^{N-1}+ \cdots + v_{iN} 2^0$. Thus, the dataset can be taken as a new vector $\overrightarrow{V}=(\bm{v_1},\bm{v_2},\cdots,\bm{v_M})$.
  We assume we have a quantum access to the vector in a quantum random access memory (QRAM)~\cite{ref23,ref24,ref25,ref26}, and the classical elements $\bm{v_i}$ ($i\in [M]$) store in $M$ memory cells. Possible physical realizations and architectures for the QRAM are discussed in detail in Ref.~\cite{ref23} and Ref.~\cite{ref25}. Then there exists an oracle $O_{V}$
  \begin{equation}
  O_{V}:\left | i \right \rangle \left | 0 \right \rangle \mapsto \left | i \right \rangle \left | \bm{v_{i}} \right \rangle,
  \label{eq:3}
  \end{equation}
  which can efficiently access $ \bm{v_{i}}$ in time $O\left(\log_{2} {M}\right)$.
  \par Before performing $O_{V}$, one important step is preparing the state $\frac{1}{\sqrt{M}}\sum_{i=1}^{M} \left | i \right \rangle$. Ref.~\cite{ref17} gives an efficient approach to generating this state by performing a product of Hadamard gates,  $H=\frac{1}{\sqrt{2}}\left ( \left(\left | 0 \right \rangle+\left | 1 \right \rangle \right)\left\langle 0 \right|+(\left | 0 \right \rangle-\left | 1 \right \rangle)\left\langle 1 \right| \right)$, and the quantum comparator that can be applied to compare the size of two numbers~\cite{ref27}. At first, prepare $m=\left \lceil \log_{2} \left( M +1 \right)\right \rceil$ qubits in $\left |0 \right \rangle$, and perform Hadamard gates to obtain  $\frac{1}{\sqrt{2^{m}}}\sum_{i=0}^{2^{m-1}} \left |i \right \rangle$. Next, we append two flag qubits. If $i=0$, flip the first qubit, and if $i>M$, flip the second qubit. The first operation can be implement by CNOT gate, and the second operation can be achieved with the help of the quantum comparator. Then measure the flag qubits, and we can obtain the state $\frac{1}{\sqrt{M}}\sum_{i=1}^{M}\left | i \right \rangle$. The measure probability is $M/2^m$, thus, the running time is $O\left(2^m/M\right)=O\left(1\right)$. Next, performing oracle $O_{V}$, we can obtain the state $\frac{1}{\sqrt{M}}\sum_{i=1}^{M}\left | i \right \rangle \left | \bm{v_{i}}\right \rangle$.
\par For the testing vector $\overrightarrow{x}$, we need prepare $N$ qubits. If $x_{j}=0$, the $j$th qubit is in the state $\left | 0 \right \rangle$. Otherwise, it is in the state $\left | 1 \right \rangle$. In this way, we have $\left | \bm{x}\right \rangle$. As a result, we can generate the state $\left | \phi _{1}\right \rangle$ in $O\left(\log_{2}{M}\right)$.
  \par A2: Implement the CNOT gate to see whether the state in the same place is equal. The CNOT gate is one of the most common controlled operations in quantum computing, with two input qubits, known as the target qubit and control qubit, respectively. The action of the $CNOT_{ct}$ gate is given by $\left | x \right \rangle_{c} \left | y \right \rangle_{t} \rightarrow \left | x \right \rangle_{c} \left |x\oplus y \right \rangle_{t}$, where $x,y \in \left \{ 0,1 \right \}$, $c$ and $t$ represent the control qubit and the target qubit respectively; that is, if $x=y$, then the target qubit $t$ is in the state $\left | 0 \right \rangle$, otherwise $t$ is in the state $\left | 1 \right \rangle$. By implementing CNOT gates on the qubits in the same place of $\left | \bm{x}\right \rangle$ and $\left | \bm{v_{i}}\right \rangle$, and labeling the result state of the target qubit $\left | x_{i} \right \rangle$ with $\left | r_{ij} \right \rangle$, we have
  \begin{align}
  \left | \phi _{2}\right \rangle&=\frac{1}{\sqrt{M}}\sum_{i=1}^{M}\bigotimes_{j=1}^{N}CNOT_{c_{ij}t_{j}}  \left | i \right \rangle \left | \bm{v_{i}}\right \rangle_{c_{i1}\cdots c_{iN}} \left | \bm{x} \right \rangle_{t_{1}\cdots t_{N}} \nonumber\\
  &=\frac{1}{\sqrt{M}}\sum_{i=1}^{M}\left | i \right \rangle \left | v_{i1} \cdots v_{iN} \right \rangle_{c_{i1}\cdots c_{iN}}\left | r_{i1} \cdots r_{iN}\right \rangle_{t_{1}\cdots t_{N}},
  \label{eq:4}
  \end{align}

  \par
\begin{figure}
  \centering
  \subfigure[]{
  \label{fig2.subfig:a}
  \begin{minipage}[a]{0.45\textwidth}
  \begin{center}
  \includegraphics[width=1\textwidth]{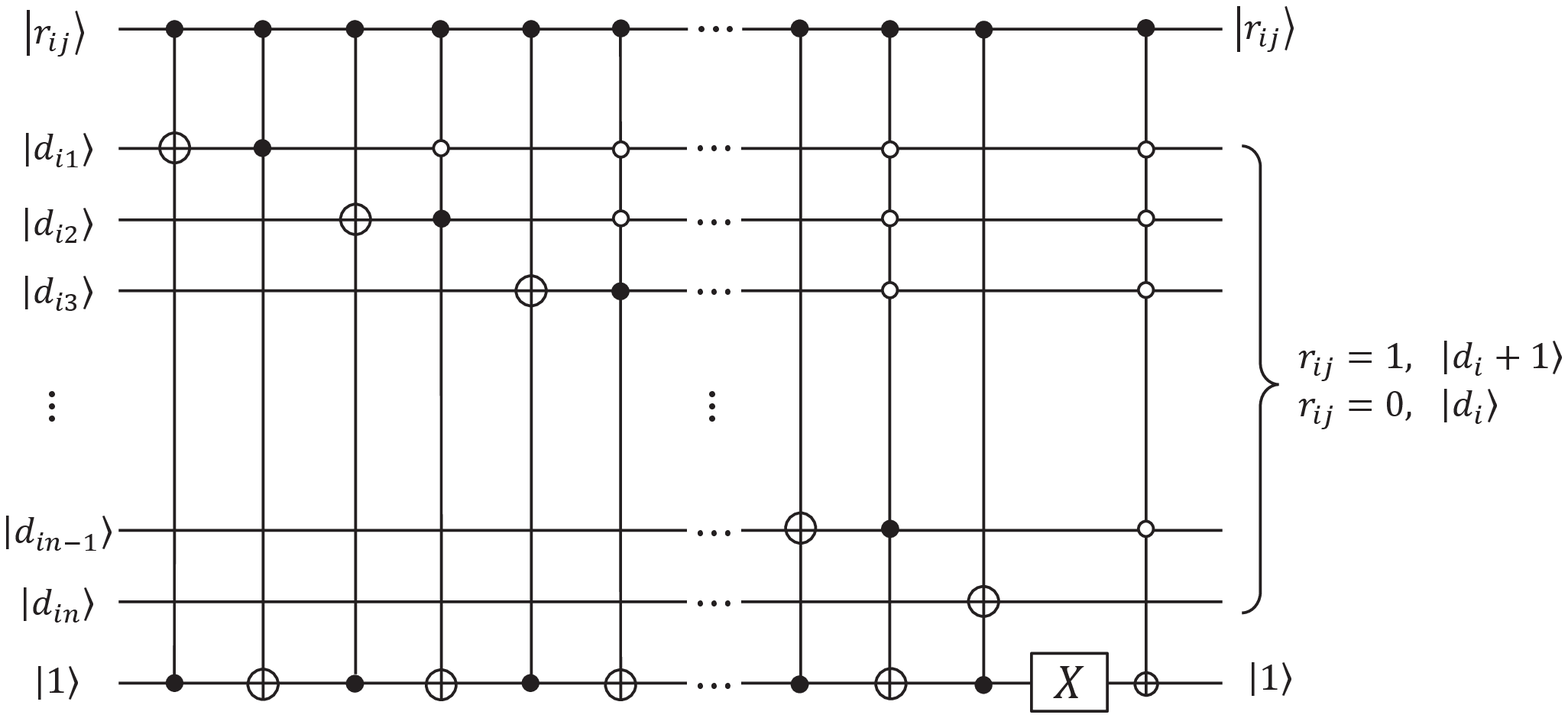}
  \end{center}
  \end{minipage}
  }
  \subfigure[]{
  \label{fig2.subfig:b}
  \begin{minipage}[b]{0.45\textwidth}
  \begin{center}
  \includegraphics[width=0.35\textwidth]{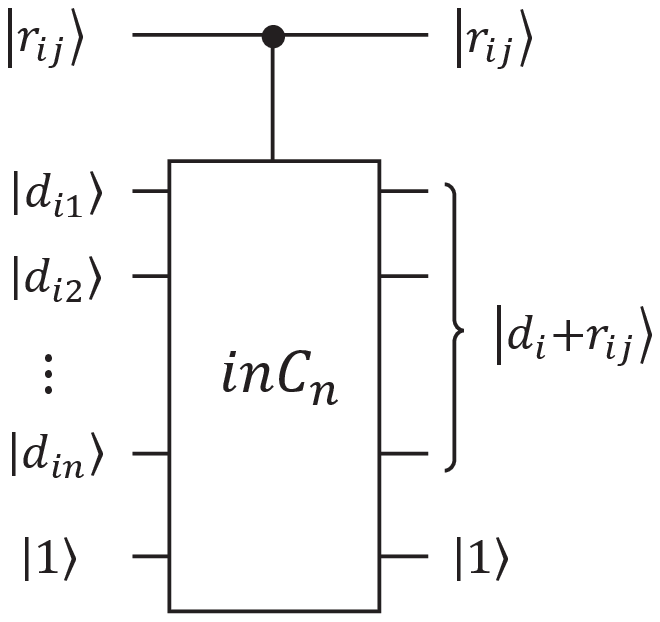}
  \end{center}
  \end{minipage}
  }
  \caption{The $d_{i}+r_{ij}$ quantum circuit. The hollow circle on the control qubit indicates that the operation is applied to the target qubit conditioned on the control qubit equalling 0. While, the solid circle is in opposite condition.}
  \label{Fig:2}
  \end{figure}

 A3: Add a register in the state $\left|0\right\rangle$ with $n=\left \lceil \log_{2} \left( N +1 \right)\right \rceil$ qubits to store Hamming distance $d_{i}$. After calculating $\sum_{j=1}^{N}  r_{ij}$, the state becomes
  \begin{equation}
      \left | \phi _{3}\right \rangle=\frac{1}{\sqrt{M}}\sum_{i=1}^{M}\left | i \right \rangle \left | v_{i1} \cdots v_{iN} \right \rangle\left | r_{i1} \cdots r_{iN}\right \rangle\left | d_{i}\right \rangle.
      \label{eq:5}
  \end{equation}
  This step can be achieved by using $d_{i}+r_{ij}$ quantum circuit proposed by Kaye\cite{ref28} as shown in Fig.~\ref{fig2.subfig:a}. In Fig.~\ref{fig2.subfig:a}, $d_{i}$ is described as $d_{i1}d_{i2} \cdots d_{in}$ in binary representation, and the controlled circuit consists of a series of controlled $NOT$ operations and a Pauli operation $X=\left|0\right\rangle \left\langle 1 \right|+\left|1\right\rangle \left\langle 0 \right| $, where controlled $NOT$ operation requires more than one control qubits. It's easy to verify that this circuit can obtain the result of $d_{i}+r_{ij}$. Moreover, this circuit can be taken as a controlled operation $inC_{n}$, as shown in Fig. \ref{fig2.subfig:b}.
  \par
\begin{figure}
  \centering
  \includegraphics[width=0.45\textwidth]{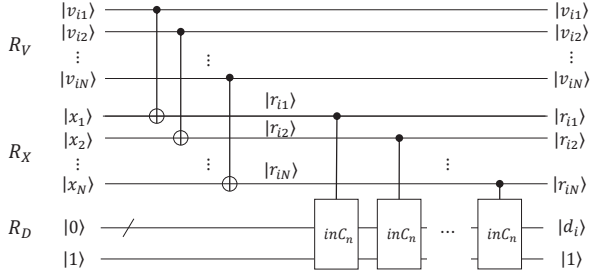}
  \caption{The circuit for computing Hamming distance $d_{i}$. The last qubit in the state $\left |1 \right \rangle$ is used for the implementation of $inC_{n}$ operation.}
  \label{Fig:3}
  \end{figure}

 Fig.~\ref{Fig:3} shows the processes of the calculation of Hamming distance. After preparing the state $\left | \phi _{1}\right \rangle$, implement CNOT gates and we can obtain  $\left | \phi _{2}\right \rangle$. Then, perform $inC_{n}$ to obtain Hamming distance $d_i$ stored in $R_D$. An example is given for illustrating this circuit more clearly. Given two vectors, $\overrightarrow{v_{i}}=(01101)^T$ and $\overrightarrow{x}=(11001)^T$, their corresponding quantum states are $\left |  v_{i1} \cdots v_{i5}\right \rangle_{R_V} = \left | 01101\right \rangle$ and $\left | x_{1} \cdots x_{5}\right \rangle_{R_X} = \left | 11001 \right \rangle$. Perform CONT gates, and the state of register $R_X$ becomes  $\left | r_{i1} \cdots r_{i5}\right \rangle_{R_X} = \left | 10100 \right \rangle$. Then, implement the operation $inC_3$ to obtain Hamming distance, $\sum_{j=1}^{5} r_{ij}$. Here, $inC_3$ needs to be performed 5 times. For the first, the first qubit of the register $R_X$ is the control qubit, and after performing the operation $inC_3$, the state of the register $R_D$ becomes $\left |0+r_{i1}\right\rangle=\left |1\right\rangle$. For the second, perform the operation $inC_3$ controlled by the second qubit of the register $R_X$, and the state of the register $R_D$ becomes $\left |1+r_{i2}\right\rangle=\left |1\right\rangle$. The last three operations are similar to the first two. As a result, we have the Hamming distance between $\overrightarrow{v_{i}}$ and $\overrightarrow{x}$, $\left |d_i\right\rangle=\left |2\right\rangle$.


\subsection{  \label{subsec:3.2}Quantum method for searching the minimum of unordered integer sequence}

\par  The second step of KNN classification algorithm is finding out $K$ nearest neighbors, that is, searching the $K$ minimum Hamming distances between the testing sample and training samples. In other words, this task can be depicted as a problem: searching the $K$ minimum elements of the set $D=\left \{d_{1},d_{2}, \cdots,d_{M}\right \}$, where $d_{i}\in \left \{ 1, 2,\cdots ,N \right \}$. Here, because the case where the distance is equal to 0 is trivial and easy to solve, we don't take it into account in this section, and the solution of that is presented in Sec.~\ref{sec:4}. To solve this problem, we propose a quantum algorithm for searching the minimum value of unordered integer sequence with runtime $O\left (  \sqrt{M}\log_{2} M  \right )$. Compared with the classical algorithm, our algorithm can achieve quadratical speedup.
\par To find out the minimum element of $D$, we introduce a parameter $key_{j}\in \left \{ 1, 2,\cdots ,N \right \}$ with its binary representation $key_{j1} key_{j2} \cdots key_{jn}$, where $n=\left\lceil \log_{2}{\left (N +1 \right )} \right\rceil$. The binary representation of $d_{i}$ is $d_{i1}d_{i2} \cdots d_{in}$. Add a bit in 0 before $d_{i}$ and $key_{j}$, i.e., $d_{i}=0d_{i1}d_{i2} \cdots d_{in}$ and $key_{j}=0key_{j1}key_{j2} \cdots key_{jn}$. Then we have $d_{i}\circleddash key_{j} =d_{i}-key$ mod $2^{n+1}$. Suppose the result is $b_{i}$ with the binary representation $b_{i0}b_{i1} \cdots b_{in}$. If $d_{i} < key$, we can obtain $b_{i0}=1$, and $b_{i1} \cdots b_{in}$ is the binary representation of $d_{i}-key_{j}$ mod $2^n$. That is, if $d_{i} < key_{j}$, the bit $b_{i0}$ will be flipped to 1, otherwise $b_{i0}=0$. This idea inspires us that the elements less than $key_{j}$ can be reserved by measuring an ancillary qubit with outcome 1, if $d_{i}$ and $key_{j}$ are encoded to quantum represents $\left | d _{i} \right \rangle$ and $\left | key_{j} \right \rangle$. What's more, thanks to quantum parallelism,  $\left | d_{i} - key\right \rangle$ can be computed in parallel, which makes the process more efficiently. Based on these analysis, a quantum algorithm for searching the minimum value of unordered integer sequence is given as follows.

  \par B1: Prepare the quantum state
  \begin{equation}
  \left | \alpha _{1} \right \rangle = \frac{1}{\sqrt{M}}\sum_{i=1}^{M}\left | i \right \rangle \left | d_{i} \right \rangle_{R_{D}},
  \label{eq:6}
  \end{equation}
  For easy of depiction, $R_{D}$ is used to denote the register storing $d_{i}$ which has $n$ qubits.
  \par \hspace{1 em} To prepare $\left | \alpha _{1} \right \rangle $, assume we are provided the quantum oracle
  \begin{equation}
  O_{D}:\left | i \right \rangle \left | 0 \right \rangle \mapsto \left | i \right \rangle \left | d_{i} \right \rangle,
  \label{eq:7}
  \end{equation}
  which can efficiently access the entries of $D$ in $O\left(\log_{2}{M}\right)$ time. This holds when the entries of $D$ are efficiently computable or are stored in QRAM~\cite{ref23,ref24,ref25,ref26}. We start with performing the oracle $O_{D}$ on the state $\frac{1}{\sqrt{M}}\sum_{i=1}^{M}\left| i \right\rangle\left| 0 \right\rangle$ to have $\frac{1}{\sqrt{M}}\sum_{i=1}^{M}\left| i \right\rangle\left| d_{i} \right\rangle $. Here, the preparation of the state $\frac{1}{\sqrt{M}}\sum_{i=1}^{M}\left| i \right\rangle$ is shown clearly in step~A1 of Sec.~\ref{subsec:3.1}.

 \par B2: Add the second ancillary qubit $a_{2}$ in the state $\left | 0 \right \rangle$ before $R_{D}$, and we have
  \begin{equation}
  \left | \alpha _{2} \right \rangle = \frac{1}{\sqrt{M}}\sum_{i=1}^{M}\left | i \right \rangle \left | 0 \right \rangle_{a_{2}}\left | d_{i} \right \rangle_{R_{D}}.
  \label{eq:9}
  \end{equation}

 \par B3: Append a register storing $key$ denoted by $R_{k}$. At first, suppose $max_{0}= N$, $min_{0}=1$, then $key_{0}=\left \lfloor \frac{max_{0}+min_{0}}{2} \right \rfloor$ with the binary representation $key_{0}=key_{01}key_{02}\cdots key_{0n}$. Generate state $\left | key _{0} \right \rangle=\left |0key_{01}key_{02}\cdots key_{0n} \right \rangle$ stored in $R_{K}$, and obtain
      \begin{equation}
      \left | \alpha _{3} \right \rangle = \frac{1}{\sqrt{M}}\sum_{i=1}^{M}\left | i \right\rangle {\left | 0 \right\rangle}_{a_{2}} {\left | d_{i} \right \rangle }_{R_{D}} {\left | key _{0} \right \rangle}_{R_{K}}.
      \label{eq:10}
      \end{equation}

 \par B4: Reserve the elements less than $key_{0}$.

  Calculate $\left | d_{i}-key_{0} \right \rangle$ in parallel, and the state becomes
  \begin{equation}
  \left | \alpha _{4} \right \rangle = \frac{1}{\sqrt{M}}\sum_{i=1}^{M}\left | i \right \rangle {\left | d_{i}\circleddash key_{0} \right \rangle} _{a_{2},R_{D}} \left | key _{0} \right \rangle _{R_{K}},
  \label{eq:11}
  \end{equation}
  where, $a \circleddash b$ represents $a - b$ mod $2^{n+1}$.
  \par \hspace{1 em} If there exist elements less than $key_{0}$ in sequence $D$, labeled with $d_{i_{p}}$, then $a_{2}:\left | 0 \right \rangle\rightarrow\left | 1\right \rangle $. We have
  \begin{align}
  \left | \alpha _{4} \right \rangle = &\frac{1}{\sqrt{M}}( \sum_{p=1}^{l} \left | i_{p}\right \rangle \left | 1\right \rangle _{a_{2}} \left | d_{i_{p}}- key_{0}\right \rangle_{R_{D}}\left | key_{0}\right \rangle_{R_{K}} \nonumber\\
  &+\sum_{q=1}^{M-l} \left | i_{q}\right \rangle \left | 0\right \rangle _{a_{2}} \left | d_{i_{q}}- key_{0}\right \rangle_{R_{D}} \left | key_{0}\right \rangle_{R_{K}}),
  \label{eq:12}
  \end{align}
  where, $l$ is the number of the elements less than $key_{0}$; and $d_{i_{q}}$ is the element bigger than or equal to $key_{0}$.
  \par\hspace{1 em} Otherwise,
  \begin{equation}
  \left | \alpha _{4} \right \rangle = \frac{1}{\sqrt{M}}\sum_{q=1}^{M}\left | i_{q} \right \rangle \left | 0 \right \rangle_{a_{2}}\left | d_{i_{q}}- key_{0} \right \rangle_{R_{D}} \left | key_{0} \right \rangle_{R_{K}}.
  \label{eq:13}
  \end{equation}
  \par \hspace{1 em}At this time, if measure $a_{2}$ to see the outcome $1$, after successful measurement, we retain the elements less than $key_{0}$. According to Eq.~(\ref{eq:12}), the probability of obtaining measurement outcome $1$ is $Prob_{2}\left (1 \right)=l/M$. The state after measurement is
  \begin{equation}
  \left | \alpha _{5} \right \rangle = \frac{1}{\sqrt{l}} \sum_{p=1}^{l}  \left | i_{p}\right \rangle \left | 1\right \rangle_{a_{2}} \left | d_{i_{p}}- key_{0}\right \rangle_{R_{D}} \left | key_{0}\right \rangle_{R_{K}}.
  \label{eq:14}
  \end{equation}
  Here, we need $O\left( M \right)$ measurements. By utilizing amplitude amplification~\cite{ref29}, the runtime can be speed up to $O\left(\sqrt{M}\right)$. If all the measurement outcomes are $0$, then the conclusion is that no element is less than $key_{0}$.

 \par Next, restore $R_{D}$ to $d_i$. Perform quantum addition operation, and we obtain
  \begin{equation}
  \left | \alpha _{6} \right \rangle = \frac{1}{\sqrt{l}} \sum_{p=1}^{l}  \left | i_{p}\right \rangle \left | 0\right \rangle_{a_{2}} \left | d_{i_{p}}\right \rangle_{R_{D}} \left | key_{0}\right \rangle_{R_{K}},
  \label{eq:15}
  \end{equation}
  or,
    \begin{equation}
  \left | \alpha _{6} \right \rangle = \frac{1}{\sqrt{M}}\sum_{q=1}^{M}\left | i_{q} \right \rangle \left | 0 \right \rangle_{a_{2}}\left | d_{i_{q}} \right \rangle_{R_{D}} \left | key_{0} \right \rangle_{R_{K}}.
  \label{eq:16}
  \end{equation}

  \par After implementing step B4, we obtain the elements less than $key_{0}$ if they exist.

  \par
\begin{figure}[h]
    \centering
    \includegraphics[width=0.5\textwidth] {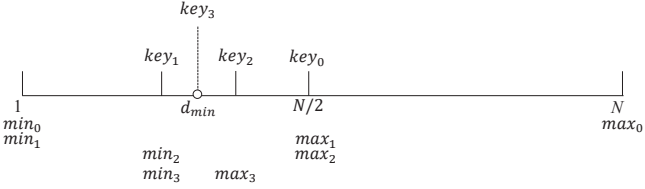}
    \caption{The process of updating the value of $key$, where $d_{min}$ denotes the minimum value of set $D$, and $min_j$, $max_j$, $key_j$ is the value of $min$, $max$, $key$ in $j$th iteration respectively.}
    \label{Fig:4}
 \end{figure}

 B5: Update the value of $key$ based on the idea of binary searching. If the outcome of measuring $a_{2}$ is $1$, then $max_{j}=key_{j-1}-1$, $min_{j}=min_{j-1}$. Otherwise, $min_{j}=key_{j-1}+1$, $max_{j}=max_{j-1}$. We have $key_{j}=\left \lfloor  \frac{max_{j}+min_{j}}{2}\right \rfloor$. For example, we assume that the minimum value of set $D$, $d_{min}$=$\frac{5N}{16}$, and $N$ is a multiple of 16, as shown in Fig.~\ref{Fig:4}. Then, the initial $max_0=N$, $min_0=1$, and $key_0=\left \lfloor  \frac{N+1}{2}\right \rfloor=\frac{N}{2}$. In the next three iteration, we have $key_1=\frac{N}{4}$, $key_2=\frac{3N}{8}$, $key_3=\frac{5N}{16}$, respectively. Obviously, through three iterations, $key$ is close to $d_{min}$.

 \par Then, update $R_{K}$ to store $\left | key_{j} \right \rangle$. Repeat step B4 until $max_{j}=min_{j}$. After measuring $a_{2}$, the state will be in two cases:
 \begin{align}
 &\left | \alpha _{7} \right \rangle =\frac{1}{\sqrt{E}}\sum_{e=1}^{E} \left| i_{e} \right\rangle \left | 1\right \rangle_{a_{2}}\left | d_{i_{e}}- key_{j} \right\rangle_{R_{D}} \left | key_{j} \right \rangle_{R_{K}},
 \label{eq:22}\\
 &\left | \alpha _{7} \right \rangle =\frac{1}{\sqrt{E}}\sum_{e=1}^{E} \left| i_{e} \right\rangle \left | 0\right \rangle_{a_{2}}\left | d_{i_{e}}- key_{j} \right\rangle_{R_{D}} \left | key_{j} \right \rangle_{R_{K}},
 \label{eq:23}
 \end{align}
 where, $E$ is the number of the elements whose value are the minimum; $d_{i_{1}}=d_{i_{2}}= \cdots = d_{i_{E}}=d_{min}$, $d_{min}$ is the value of the minimum of set $D$. Eq.~(\ref{eq:22}) can be obtained when the measure outcome is $1$, where $d_{min}<key_{j}$, and Eq.~(\ref{eq:23}) can be obtained when the measure outcome is $0$, where $d_{min}=key_{j}$. Then performing addition operation on register $R_{D}$ and $R_{K}$, the final state is
 \begin{equation}
  \left | \alpha _{min} \right \rangle = \frac{1}{\sqrt{E}}\sum_{e=1}^{E} \left | i_{e} \right \rangle \left | 0\right \rangle_{a_{2}}\left | d_{i_{e}} \right \rangle_{R_{D}}\left | key_{j} \right \rangle_{R_{K}}.
  \label{eq:24}
  \end{equation}
  \par Measuring the index register and $R_{D}$ in computational basis, we obtain the value of the minimum elements of $D$, $d_{min}$ and its index.

 \subsection{\label{subsec:3.3}Runtime analysis}

  \par One of the clear advantages of quantum algorithm is that it is faster than classical algorithm. Hence, in this section, the runtime of the above two quantum sub-algorithms are discussed respectively. Similarly to the representative quantum algorithm, we analyze the number of quantum operations in each step of the two sub-algorithms and obtain the time complexity of them. It is shown that the presented  sub-algorithms achieve significant speedup over their classical counterparts.
  \par The sub-algorithm A presents a quantum method for calculating Hamming distance. Obviously, in step~A1, the runtime of state preparation is $O\left(\log_{2}{M}\right)$. Then, $N$ CNOT gates are utilized to obtain $\left | r_{i1} \cdots r_{iN}\right \rangle$ in step~A2, so the time complexity of this step is $O\left(N\right)$. Finally, to compute the sum of $r_{ij}$, $N$ $inC_{n}$ operations are used in step~A3. According to Ref.~\cite{ref28}, the depth of $inC_{n}$ is
  \begin{equation}
  \left\{\begin{matrix}
  1&n=1\\
  10&n=2\\
  2n^2+n-5&n\geq 3
  \end{matrix}\right.
  \end{equation}
Thus, each $inC_n$ operation has $O\left(n^2\right)$ elementary gates, which means that step~A3 takes runtime $O\left(Nn^2\right)$. Based on the runtime of each step of sub-algorithm A shown in Table~\ref{tab:table1}, we can obtain the overall runtime of this sub-algorithm is $O\left(\log_{2}{M}+N+Nn^2\right)$. Considering a low-dimensional feature space, i.e., $N \ll  M$, we can ignore the effect of $N$ on the overall algorithm. In this case, the time complexity of the quantum sub-algorithm A is $O\left(\log_{2}{M}\right)$. As compared to the classical counterpart, the runtime of which is $O \left( M \right)$, the proposed quantum sub-algorithm greatly reduces the time complexity.
  \begin{table}[h]
  \caption{\label{tab:table1}
  The runtime of the sub-algorithm A. }
  \begin{ruledtabular}
  \begin{tabular}{cc}
  \textrm{Step}&\textrm{Runtime}\\
  \colrule
  \textrm{A1}&$O\left(\log_{2}{M}\right)$ \\
  \textrm{A2}&$O\left(N\right)$\\
  \textrm{A3}&$O\left(Nn^2\right)$\\
  \colrule
  \textrm{Overall time complexity}&$O\left(\log_{2}{M}+N+Nn^2\right)$\\
  \end{tabular}
  \end{ruledtabular}
  \end{table}

  \par The sub-algorithm B is a quantum sub-algorithm for searching the minimum of unordered integer sequence which can be used to find out the $K$ nearest neighbors in KNN classification algorithm. In step B1, QRAM~\cite{ref23,ref24,ref25,ref26} is utilized to state preparation with the runtime $O\left(\log_{2}{M}\right)$. Then, in step B2 and step B3, the particle $a_{2}$ in state $ \left | 0\right \rangle$ and the register $R_{K}$ in state $ \left | key_0\right \rangle$ are appended respectively, so the runtime of the two steps is $O \left( 1 \right)$. After that, in Step B4, $\left | d_{i}-key \right \rangle$ is calculated in parallel, and $a_{2}$ is measured to reserve the elements less than $key$. The probability of obtaining measurement outcome $1$ of $a_{2}$ is $l/M$ where $l$ is unkonwn, so it need $O\left(\sqrt{M}\right)$ repetitions by utilizing amplitude amplification~\cite{ref29}. Finally, $key$ is updated $O \left(\log_{2}{N} \right)$ times to obtain the minimum value in step~B5. Based on Table~\ref{tab:table2}, the  overall time complexity of the sub-algorithm B is $O\left(\log_{2}{N}\sqrt{M}\log_{2}{M}\right)$. Here, the case where the sample vectors lie in a low-dimensional feature space is also taken into account, i.e., $N \ll M$, the time complexity is $O\left(\sqrt{M}\log_{2}{M} \right)$. Compared with the classical searching algorithm with runtime $O\left( M \right)$, obviously, our algorithm achieves the speedup.

  \begin{table}[h]
  \caption{\label{tab:table2}%
  The runtime of sub-algorithm B.}
  \begin{ruledtabular}
  \begin{tabular}{cc}
  \textrm{Step}&\textrm{Runtime}\\
  \colrule
  \textrm{B1}&$O\left(\log_{2}{M}\right)$ \\
  \textrm{B2}&$O\left(1\right)$\\
  \textrm{B3}&$O\left(1\right)$\\
  \textrm{B4}&$O\left(\sqrt{M}\right)$\\
  \textrm{B5}&$O\left(\log_{2}{N}\right)$\\
  \colrule
  \textrm{Overall time complexity}&$O\left(\log_{2}{N}\sqrt{M}\log_{2}{M}\right)$\\
  \end{tabular}
  \end{ruledtabular}
  \end{table}

 \par Involving the problem of searching the minimum value, Christoph D\"{u}rr and Peter H\o yer proposed a quantum algorithm for finding the minimum value based on Grover's search algorithm~\cite{ref21} with running time $O\left(\sqrt{M}\right)$ ($M$ is the sample scale)  in 1996~\cite{ref22}. Given a set containing $M$ elements, their algorithm utilizes an oracle to compare the value of elements with one of them and mark the smaller. The element can be in a arbitrary numeric set. Different with their algorithm, the data set in our algorithm is required to be a set of integers in a known range. We search the minimum based on the properties of elements and we don't rely on the oracle to mark the index register. The elements less than $key$ are reserved by measuring a flag qubit, which can be implemented via the present quantum technique, such as QRAM~\cite{ref23,ref24,ref25,ref26}, quantum addition operation~\cite{ref28} and amplitude amplification~\cite{ref29}. In practical term, there exist lots of application scenarios where the element requires to be a integer and its scope has made requirement as well, such as finding the minimum students' test score in student management system, finding the minimum age in Demographic and Health Surveys and finding the minimum Hamming distance which we concern in this paper.
\section{\label{sec:4}Whole quantum frame for KNN classifier}

\par Based on the above two sub-algorithms, in this section, we present the whole quantum frame for KNN classifier based on Hamming distance in detail in Sec.~\ref{subsec:4.1}. Its runtime analysis is in Sec.~\ref{subsec:4.2}.
 \subsection{\label{subsec:4.1}Quantum algorithm}

Given testing sample vector $\overrightarrow{x}$, and training sample vectors $\{\overrightarrow{v_{i}},c_{i}\}_{i=1}^M$, the detail of the algorithm is as follows.
  \par Step W1: Prepare the initial quantum state.
  \par In this step, the initial quantum state
  \begin{equation}
  \left | \psi_{0}\right \rangle=\frac{1}{\sqrt{M}}\sum_{i=1}^{M}\left |i \right \rangle \left |v_{i1} \cdots v_{iN} \right \rangle_{R_{V}} \left | c_{i}\right \rangle_{R_{C}} \left | x_{1} \cdots x_{N}\right \rangle_{R_{X}}
  \label{eq:25}
  \end{equation}
  is generated.
  \par We assume training samples are stored in QRAM~\cite{ref23,ref24,ref25,ref26}, and oracle $O_{VC}$ is provided, where
  \begin{equation}
  O_{VC}:\left | i \right \rangle \left | 0 \right \rangle\left | 0 \right \rangle \mapsto \left | i \right \rangle \left | \bm{v_{i}}\right \rangle \left |c_{i}\right \rangle
  \label{eq:26}
  \end{equation}
  Then, the state $\frac{1}{\sqrt{M}}\sum_{i=1}^{M}\left |i \right \rangle\left |\bm{v_{i}}\right \rangle\left |c_{i}\right \rangle$ can be obtained in $O\left(\log_{2}{M}\right)$. To generate $\left | \bm{x}\right \rangle=\left | x_{1} x_{2} \cdots x_{N}\right \rangle$, prepare the state $\left |0 \cdots 0\right \rangle$ with $N$ qubits at first. If $x_{j}=1$, then the $j$th qubit is flipped to $\left | 1 \right \rangle$. Otherwise, it is still in $\left | 0 \right \rangle$. Thus, we can obtain $\left | \bm{x}\right \rangle$ stored in $R_{X}$. As a result, $\left | \psi_{0}\right \rangle$ is generated in $O\left(\log_{2}{M}\right)$.
  \par Step W2: Calculate Hamming distance.
  \par  In this step, the first sub-algorithm is utilized to compute Hamming distances between the testing sample and training samples. Based on the quantum circuit as shown in Fig.~\ref{Fig:3} to calculate Hamming distance described in Sec.~\ref{subsec:3.1}. firstly, perform CNOT gates on $R_{V}$ and $R_{X}$ to see whether the state in the same place is equal. Then, $N$ controlled operation $inC_{n}$ are performed to obtain $\left |d_{i} \right \rangle$. The state becomes
  \begin{align}
  \left | \psi_{1}\right \rangle &=  \frac{1}{\sqrt{M}}\sum_{i=1}^{M}\left |i \right \rangle \left |v_{i1} \cdots v_{iN} \right \rangle_{R_{V}}\left |c_{i}\right \rangle _{R_{C}}\left | r_{i1} \cdots r_{iN}\right \rangle_{R_{X}}\left |d_{i} \right \rangle_{R_{D}}\nonumber \\
  & =\frac{1}{\sqrt{M}}\sum_{i=1}^{M}\left |i \right \rangle \left |\bm{v_{i}} \right \rangle_{R_{V}}\left |c_{i}\right \rangle _{R_{C}}\left | \bm{r_{i}}\right \rangle_{R_{X}}\left |d_{i} \right \rangle_{R_{D}}.
  \label{eq:27}
  \end{align}

  Step W3: Search $K$ nearest neighbors.
  \par Considering the case where $d_{i}\neq 0$, $d_{i}$ is an integer between $1$ and $N$. We can use the second sub-algorithm proposed in Sec.~\ref{subsec:3.2}. to search $K$ nearest neighbors of $\overrightarrow{x}$ directly. When the case where $d_{i}=0$ is taken into account, the distance $d_{i}$ plus $1$ can be calculated in advance so that the scope becomes $\left \{ 1, 2,\cdots ,N+1 \right \}$, and the second sub-algorithm is also applicable in this case. Here, we present the process in the case where $d_{i}\neq 0$.
  At this time, by implementing the second sub-algorithm, we obtain the index $i_{min}$ and category $c_{i_{min}}$ of the nearest neighbor. Next, the steps are described in brief.
    \par Step W3.1: As shown in steps B2 and B3, append the second ancillary qubit $a_{2}$ as a flag particle in the state $\left |0 \right \rangle$, and a register $R_{K}$ to store the state $\left |key_{0} \right \rangle$, where $key_{0}=\left \lfloor \frac{max_{0}+min_{0}}{2} \right \rfloor$ with the binary representation $key_{0}=key_{01}key_{02}\cdots key_{0n}$ and initial $max_{0}= N$, $min_{0}=1$.
    \par Step W3.2: Perform step B4 to reserve the elements less than $key_{0}$. Calculate $\left | d_{i}-key_{0} \right \rangle$ at first, then measure $a_{2}$. If the outcome is $1$, the elements less than $key$ will be reserved. The probability of the outcome $1$ will be $l/M$, so to perform measurements, we need to make $O\left(\sqrt{M}\right)$ measurements. One case is that all the measurement outcomes are equal to $0$,  which shows all elements are not less than $key_0$. Restore $R_{D}$ to $d_i$,  we obtain the state
    \begin{align}
    \left | \psi_{2} \right \rangle =& \frac{1}{\sqrt{l}} \sum_{p=1}^{l} \left | i_{p}\right \rangle\left |\bm{v_{i_{p}}} \right \rangle_{R_{V}}\left |c_{i_{p}}\right \rangle _{R_{C}}\left | \bm{r_{i_{p}}}\right \rangle_{R_{X}}\left | 0\right \rangle_{a_{2}} \nonumber \\
    & \left | d_{i_{p}}\right \rangle_{R_{D}} \left | key_{0}\right \rangle_{R_{K}},
    \label{eq:29}
    \end{align}
    or
    \begin{align}
    \left | \psi_{3} \right \rangle =& \frac{1}{\sqrt{M}} \sum_{q=1}^{M}  \left | i_{q}\right \rangle\left |\bm{v_{i_{q}}} \right \rangle_{R_{V}}\left |c_{i_{q}}\right \rangle _{R_{C}}\left | \bm{r_{i_{q}}}\right \rangle_{R_{X}}\left | 0\right \rangle_{a_{2}}\nonumber \\
    & \left | d_{i_{q}}\right \rangle_{R_{D}} \left | key_{0}\right \rangle_{R_{K}}.
    \label{eq:30}
    \end{align}

    \par Step W3.3: Update the value of $key$ as step B5 shows, and repeat step W3.2 until $max_{j}=min_{j}$. Finally, after measuring the index register and  register $R_{C}$, $i_{min}$ and $c_{i_{min}}$ can be obtained.
    \par Step W3.4: Take $\overrightarrow{v_{i_{min}}}$ out of the training sets, and repeat the second sub-algorithm, then we can obtain the second nearest neighbor. Therefore, repeating the second sub-algorithm $K$ times, we can obtain the $K$ nearest neighbors with their index $k_{i}$ ($i\in \left\{1,\cdots ,K \right\}$) and category $c_{k_{i}}$.

    \par Step W4: Determine the testing sample's category.
    \par The task of this step is to find out the category whose frequency is the highest of the $K$ nearest neighbors. Here, the quantum measurement is utilized to avoid quantum-classical interaction, so that the whole algorithm is more efficient. Firstly, based on the result we obtain in step W3, we prepare the state,
    \begin{equation}
    \left | \theta  \right \rangle=\frac{1}{\sqrt{K}}\sum_{i=1}^{K}\left |i\right \rangle \left | c_{k_{i}} \right \rangle,
    \label{eq:31}
    \end{equation}
    where, $c_{k_{i}}$ is category of the $k_{i}$th nearest neighbor. Suppose $c_{k_{i}}$ is stored in QRAM~\cite{ref23,ref24,ref25,ref26}, and the oracle $O_{C_{k}}$
    \begin{equation}
    O_{C_{k}}:\left | i \right \rangle \left | 0 \right \rangle \mapsto \left | i \right \rangle \left | c_{k_{i}} \right \rangle
   \label{eq:32}
   \end{equation}
    is provided. After generating $\frac{1}{\sqrt{K}}\sum_{i=1}^{K}\left |i\right \rangle\left |0\right \rangle$, we can implement the oracle $O_{C_{k}}$ and obtain the state $\left | \theta  \right \rangle$ in time $O\left(\log_2{K}\right)$.
    \par Assume the number of category $j$ in $K$ nearest neighbors is $\rho_{j}$, $j\in \left\{0,1,\cdots ,L \right\}$. Measuring the second register $R_{C}$ in computational basis, the probability of outcome $j$ is
    \begin{equation}
    Prob \left(j \right)=\left \langle \theta   \right | \left(I \otimes \left | j \right \rangle\left \langle j \right |  \right)\left | \theta  \right \rangle=\frac{\rho_{j}}{K}.
    \label{eq:33}
    \end{equation}
    \par  It's easy to see that the higher the frequency of category is, the bigger measurement probability is. Besides, $O\left(K\right)$ repetitions are enough to obtain the category whose frequency is highest in $K$ nearest neighbors. At the end, assign the most frequent category to the testing sample.
 \subsection{\label{subsec:4.2}Runtime analysis}

\par In this section, the time complexity of the proposed quantum KNN classification algorithm is analyzed briefly. Based on the results of Sec.~\ref{subsec:3.3}, we analyze the runtime of each step and obtain the overall time complexity of the whole algorithm is $O\left(\sqrt{M}\log_{2}{M} \right)$. As shown in Table~\ref{tab:table3}, a detailed analysis of each step of the whole quantum algorithm is depicted as follows.

\par At first, in step~W1, the state $\left | \psi_{0}\right \rangle$ is generated in time $O \left(\log_{2}{M}\right)$ with the help of QRAM~\cite{ref23,ref24,ref25,ref26}. Then, in step~W2, the Hamming distance between testing sample and training samples are calculated in parallel by utilizing the sub-algorithm A, where $N$ CNOT gates and $N$ $inC_n$ operations are implemented. Thus, the runtime of step W2 is $O \left( N+Nn^2 \right)$. Next, sub-algorithm B is utilized to search $K$ nearest neighbors in step~W3, which is the most time consuming. The substeps of step~W3 have the corresponding steps of sub-algorithm B, where step~W3.1 corresponds step B2 and B3; step~W3.2 and step~W3.3 correspond step~B4 and step~B5 respectively. According to the runtime analysis of sub-algorithm B in Sec.~\ref{subsec:3.3}, steps B2 to B5 cost $O\left(\log_{2}{N}\sqrt{M}\right)$ in total. Besides, the sub-algorithm B is required to perform $K$ times to search the $K$ minimum distances, so the runtime on step W3 is $O\left(K\log_{2}{N}\sqrt{M}\right)$. Finally, the testing sample's category is determined in step~W4. As generating the state $\left | \theta  \right \rangle$ costs $O\left(\log_2{M}\right)$ and $O\left(K\right)$ measurements are taken, the runtime on step~W4 is $O\left(K\log_2{K}\right)$.
\par To sum up, the overall time complexity of the presented quantum KNN classification algorithm is $O\left(K^2\log_{2}{K}\log_{2}{N}\sqrt{M}\left(\log_{2}{M}+N+Nn^2\right)\right)$. Generally speaking, the vector dimension $N$ and the number of categories $K$ are always far less than the sample scale $M$, so they have a trivial impact on the overall runtime. It means that the overall time complexity of the quantum KNN classification algorithm is  $O\left(\sqrt{M}\log_{2}{M} \right)$. Compared with the classical KNN classification algorithm whose time complexity is $O\left( M \right)$, our algorithm achieves quadratical speedup when the sample size is large.
\begin{table}[h]
  \caption{\label{tab:table3}%
  The runtime of the quantum KNN classification algorithm.}
  \begin{ruledtabular}
  \begin{tabular}{cc}
  {Step}& Runtime \\
  \colrule
  W1 &  $O\left(\log_2{M}\right)$ \\
  W2 & $O\left(N+Nn^2\right)$ \\
  W3 & $O\left( K\log_2{N}\sqrt{M}\right)$ \\
  W4 &   $O\left(K\log_2{K}\right)$ \\
  \colrule
\multicolumn{2}{c} {In total: $O\left(K^2\log_2{K}\log_2{N}\sqrt{M}\left(\log_2{M}+N+Nn^2\right)\right)$}
  \end{tabular}
  \end{ruledtabular}
\end{table}

\section{\label{sec:5}Conclusions}

\par To sum up, in this paper, we consider quantum algorithm for KNN classifier based on Hamming distance, which is one of the most basic algorithms in machine learning and can be a significant subprocess in lots of classification algorithms~\cite{ref18,ref19} as well. To achieve this task, two core sub-algorithms are proposed firstly. One is quantum method to calculate Hamming distances between testing sample and training samples, where the addition circuit presented by Kaye~\cite{ref28} is utilized. Another is the sub-algorithm for searching the minimum of unorder integer sequence aiming at finding out the nearest neighbor, which also can be efficiently applied to solve some statistical problems. Based on the above methods, we put forward the whole quantum frame of KNN classification algorithm.
\par Through a brief analysis, the presented algorithm can classify the testing sample with the time complexity $O\left(\sqrt{M}\log_{2}{M} \right)$ when the sample vectors lie in a low-dimensional feature space. Thanks to the characteristics of quantum computation, the algorithm achieves quadratical speedup over the classical algorithm. However, when the dimension of sample vectors is large, it is hard to obtain the quadratical acceleration, which is a new issue we will study in the future. Moreover, how to efficiently select $K$ in quantum KNN classification algorithm also deserves further investigation.

\begin{acknowledgments}
This work was supported by National Natural Science Foundation of China (Grants No. 61976053 and No. 61772134), Fujian Province Natural Science Foundation (Grant No. 2018J01776), and Program for New Century Excellent Talents in Fujian Province University.
\end{acknowledgments}

\end{document}